\documentclass[12pt,reqno]{amsart}
\usepackage{amssymb,graphics}
\usepackage[cp1251]{inputenc}
\usepackage[russian]{babel}
\newcommand{\mfrac}[2]
{\raisebox{0.045em}{\mbox{\footnotesize$\displaystyle
\frac{#1}{#2}$}}}
\newcommand{\pot}{\mbox{\footnotesize $[$} u \mbox{\footnotesize $]$}}

\newcommand{\reff}[1]{$($\ref{#1}$)$}

\begin{document}
\centerline{{\bf Исторические замечания к теории конечнозонного интегрирования:}}
\centerline{\bf элементарная трактовка теории}

\bigskip
\centerline{\em Ю.\,В.\,Брежнев}
\bigskip\bigskip

\noindent
\section{Введение}
\noindent
Современная история конечнозонного интегрирования
насчитывает уже  30 лет.
Устоялось два названия для данной теории:
{\em конечнозонное\/}  и {\em алгебро-геометрическое интегрирование\/}. Исторически, оба связаны с интегрированием
спектральной задачи, определяемой обыкновенным дифференциальным уравнением
\begin{equation}
\Psi''-u(x)\,\Psi=\lambda\,\Psi
\end{equation}
и  знаменитым дифференциальным уравнением в частных производных --- уравнением Кортевега--де-Фриза
\begin{equation}
u_t=u_{\mathit{xxx}}-6\,u\,u_x.
\end{equation}
Ни одному уравнению не посвящено столько современной математической литературы, сколько ему и уравнению (1). 
В самом деле, с ними связано
открытие в 1967 году метода обратной задачи рассеяния  ({\sc мозр}), а последующие ответвления теории столь многочисленны, что представляют собой
безусловно самостоятельные широкие направления, нежели частные случаи. 

Термин <<конечнозонное интегрирование>>
отражает современно-исторические корни: проблема интегрирования
спектральной задачи (1). В широком смысле слова это означает
поиск <<хороших>> потенциалов $u(x)$, соответствующих решений для
$\Psi$-функции, определение дискретного и непрерывного
спектра для них и нахождение других аттрибутов классической
спектральной теории. Выяснилось, что такими потенциалами должны быть
периодические (или квази-) по $x$, а их частным случаем, когда период стремится к бесконечности, являются знаменитые солитоны в {\sc мозр}.

Второе название отражает то понимание теории, которое было достигнуто  в работах Кричевера 1970-х годов
спустя несколько лет после появления работы Новикова \cite{novikov}. Оно вскрыло весьма глубокую связь теории, носившую сначала полностью <<спектральный>> характер
(как и {\sc мозр}), с методами алгебраической геометрии и алгебраической теории дифференциальных операторов.

Итогом первоначального развития теории стало понимание того факта, что окончательные ответы всегда выражаются в терминах многомерных $\Theta$-функций, ассоциированных с алгебраическими кривыми
конечного рода. В связи с этим иногда употребляется название
{\em $\Theta$-функциональное интегрирование\/}.

Как стало ясно позднее, математический аппарат теории и даже
многие аттрибуты самой теории, встречаются задолго до
1970-х годов и связаны с именами Ковалевской, Вейерштрасса,
Клебша, Гордана, Бёрчналла--Чонди, Бейкера и других. В 1961 году Ахиезер опубликовал работу по спектральной теории  уравнения (1), на которую следуют ссылки во всех более менее обстоятельных работах по данной тематике.

\subsection{Мотивация}
В данной заметке мы хотели бы изложить тот аспект теории,
который не только исторически возник до появления упомянутых выше, но
и отражает в большей степени понимание предмета
на весьма простом уровне мотивации. Довольно неожиданно то, что
этот факт  не нашел отражения в современной  литературе, хотя имя Ж.\,Драш, которому принадлежит
идеология и центральные результаты, было обнаружено уже в начале 1980-х годов. Сам Драш опубликовал на эту тему в 1919 году  лишь две короткие заметки в Comptes Rendus без доказательств и объяснений. В связи с этим мы расширим пояснения, а необходимые дополнительные
доказательства и обоснования довольно просты и могут быть легко скомпенсированы обращением к соответствующей литературе. В приложении
мы приводим перевод двух заметок Драша в силу их чрезвычайной
краткости.
Мы указываем также на тот факт, что теория эквивалентна
другой концепции квадратурной интегрируемости, начавшейся разрабатываться Лиувиллем в 1830-х годах еще до появления в 1840--50-х годах 
знаменитой интегрируемости нелинейных гамильтоновых систем.

Излагая теорию, мы  исключаем современные аксиоматические построения и не прибегаем к  аппарату алгебраической геометрии, римановых поверхностей, 
функциям Бейкера--Ахиезера, теории операторов, их спектрам и т.\,д.
Как станет видно, <<наивная>> постановка проблемы приводит к тому, что
все объекты теории с неизбежностью появляются сами собой. Среди них: формула для $\Psi$-функции,
уравнения Новикова, алгебраическая гиперэллиптическая кривая и т.\,д. Одновременно мы получаем ответы на вопросы типа: откуда берется  фундаментальный полином по спектральному параметру $\lambda$ и переменные $\gamma_k^{}$ как нули этого полинома,  какова роль формулы следов, почему возникает проблема обращения абелевых интегралов (задача Якоби)? Несмотря на развитость теории мы приводим факт, отсутствующий в литературе: окончательную квадратурную формулу для $\Psi$-функции (формула \reff{final}). Эта формула похожа на редко упоминаемую
формулу (4.12--13) в работе \cite{matveev}, но отличается тем
чем отличаются спектральный и вышеупомянутый подходы. 
Различие  в подходах и их эквивалентность объясняет известная теорема Вейерштрасса о перестановке параметров и пределов в нормализованных абелевых интегралах 3-го рода.

Методика в действительности переносима на другие спектральные задачи и нелинейные уравнения.
На этих сравнениях и обобщениях мы не будем останавливаться, но приводим рассуждения так чтобы они с наименьшими изменениями переносились
на другие <<конечнозонные задачи>>.

В заключении, укажем что
общепринятый взгляд на $\Theta$-функциональное решение
проблемы тоже выводим, как  собственно и формула для $\Theta$-ряда или появление $\Theta$-функции Римана. Эти аспекты теории в данной заметке на затрагиваются.

\section{Краткое изложение современных результатов}
\noindent
Перечислим основные результаты теории.
Пионерская работа Новикова \cite{novikov} была нацелена на решение следующей проблемы: как перенести {\sc мозр} для (1) на случай периодических потенциалов и как выглядят, в этом случае,  аналоги всех объектов  спектральной теории, уже весьма развитой к тому времени для достаточно быстро убывающих потенциалов.
Эти объекты были построены, а сами потенциалы оказались удовлетворяющими обыкновенным дифференциальным уравнениям по переменной $x$. Эти уравнения, именуемые сейчас уравнениями Новикова, являются не чем иным как стационарными (т.\,е. когда $u_t=0$)
уравнением КдФ и высшими членами его иерархии. Со спектральной точки зрения
эти потенциалы оказались хорошими  в том смысле, что имели
конечное число <<разрешенных>> зон в своем непрерывном спектре: на вещественной оси переменной $\lambda$ имеется конечное число
интервалов, таких что если $\lambda$ берется оттуда, то
решение $\Psi(x;\lambda)$ дифференциального уравнения (1) при данном $u(x)$ является ограниченной функцией на всей оси $x\in (-\infty\ldots\infty)$. Исторически, такие задачи уже рассматривались (теория Флоке), где $x$ играло роль  времени $t$. Помимо такой серьезной физической мотивации, такие потенциалы
были хороши тем, что вся теория становилась эффективной с аналитической точки зрения, т.\,к. описывалась имеющимся и развитым математическим аппаратом. Он использует теорию алгебраических функций, их римановых поверхностей, абелевых интегралов на них и другое. Сразу обнаружилось, что теория требует серьезных доработок и расширения на квазипериодические случаи и комплекснозначные потенциалы.

Другим, не менее ценным, результатом этой работы стало осознание
того факта, что уравнения Новикова представимы как классические гамильтоновы динамические системы (по $x$). Новиков продемонстировал такую интегрируемость для случая двух зон, а позднее было доказано что это есть общий признак такой интегрируемости
(Гельфанд--Дикий). В 1975 году Дубровин и Новиков привели первый явный нетривиальный пример, когда потенциал зависит от дополнительной переменной $t$, а вся теория сохраняет вид.
Сейчас это называется изоспектральными деформациями
интегрируемых потенциалов. Мы будем употреблять этот термин применительно к потенциалу, поскольку им полностью определяется решение $\Psi$ и спектр.

Центральным объектом теории стало алгебраическое уравнение $W(\lambda,\mu)=0$, связывающее переменную $\lambda$ и собственное значение $\mu$ второго оператора $\widehat A$, коммутирующего с первым оператором $\widehat L=\frac{d^2}{dx^2}-u(x)$ и определяющего задачу (1). Этот второй оператор и его порядок зависит от того,
какой потенциал взят. Например, если мы имеем дело с самим стационарным уравнением КдФ $u_{\mathit{xxx}}-6\,u\,u_x=0$, то
оператор $\widehat A$ имеет вид
$$
\widehat A=\frac{d^3}{dx^3}-\frac32\,u\,\frac{d}{dx}-u_x.
$$
Центральная роль алгебраического соотношения $W(\lambda,\mu)=0$
заключается в том, что оно представляет собой не что иное как
наполовину проинтегрированное само стационарного уравнение: его коэффициентами являются дифференциальные полиномы по $u(x)$, но все они константы. В упомянутом 1-зонном случае мы имеем (опуская несущественные параметры) соотношение
\begin{equation}\label{g1}
W(\lambda,\mu):\;\;
\mu^2=\lambda^3+\frac14\,\big(u_\mathit{xx}-3\,u^2\big)\,\lambda-
\frac18\,u\,u_\mathit{xx}+\frac{1}{16}\,u_x^2+\frac14\,u^3,
\end{equation}
в котором все коэффициенты перед $\lambda^3,\,\lambda^1,\,\lambda^0$ есть константы интегрирования $A_k$.
Новиков показал как перейти от дифференциальных  переменных
$u,\,u_x,\ldots$ к каноническим переменным $\{p,q\}$, используемым в теории динамических систем.

Поскольку упомянутые два оператора определяют спектральные задачи
$\widehat L\Psi=\lambda\,\Psi$ и $\widehat A\Psi=\mu\,\Psi$, то они сами коммутируют, имеют общую собственную функцию $\Psi(x;\lambda)$ и связаны тем же самым алгебраическим соотношением $W(\widehat L,\widehat A)=0$. Все что сделано для оператора Шредингера $\widehat L$ будет справедливо и для $\widehat A$. 
Этот факт выявил глубокую связь теории с алгебраической теорией операторов, и как
обнаружил вскоре И.\,Кричевер \cite{krichever}, уже имел место в классической литературе, забытой к 1970-м годам (Бёрчналл--Чонди и Бейкер 1920--30-е годы). Позднее, им же, такой взгляд на уравнения был
систематизирован и приобрел сегодняшний вид. Более того, стало ясно, что теория не ограничивается уравнением Шредингера (1) и КдФ (2), переносится на более сложные операторы высоких порядков,
уравнения, матричные задачи и т.\,д. Рассматривая  соотношение $W(\lambda,\mu)=0$ как алгебраическую кривую, связывающую две переменные $\lambda$ и $\mu$, мы делаем наблюдение, что она, как риманова поверхность, всегда имеет конечный род, так как степень полинома $W(\lambda,\mu)$ по этим переменным конечна. Фактически, род этой кривой и число зон в спектре операторов $\widehat L,\,\widehat A$
есть совпадающие величины.

Как описывать данные потенциалы и решения аналитическими формулами? Это было сделано в 1975 году Матвеевым  и Итсом 
\cite{matveev} и известно как знаменитые $\Theta$-функ\-ци\-о\-наль\-ные формулы теории
конечнозонного интегрирования. Мы затронем этот вопрос позднее.

\section{Интегрируемости по Лиувиллю}
\noindent
В современных теориях интегрирования обыкновенных
дифференциальных уравнений существует две разновидности понятия интегрируемости и обе, исторически, связаны с именем Лиувилля.

Первая, и самая знаменитая, касается интегрирования {\em нелинейных\/} обыкновенных дифференциальных уравнений (систем), если они представимы в гамильтоновой форме и имеют интегралы движения в количестве равном размерности конфигурационного пространства. В другой терминологии эта интегрируемость известна
как {\em интегрируемость в квадратурах\/}, хотя непосредственная реализация квадратур представляет собой самостоятельную проблему.
Если интегралы алгебраичны, то это --- теория разделения переменных, восходящая к Якоби. Такую интегрируемость иногда называют алгебраической.

Вторая относится к периоду между 1833 и  1841 годами и, таким образом, предшествовала первой, относящейся к 1840--50-м годам.
У Лиувилля обе интегрируемости были интегрируемостями в квадратурах. Различными были лишь объекты исследования:
нелинейные гамильтоновы уравнения с одной стороны и линейные дифференциальные уравнения и вычисление неопределенных интегралов с  другой. 
Более поздняя интегрируемость нелинейных гамильтоновых систем
носила больше характер существования интегралов
(точный дифференциал $dS=p\,dq-H\,dt$), хотя Лиувилль уже был знаком к тому времени с формализмом Гамильтона--Якоби%
\footnote{Якоби был в Париже в 1842 перед тем как дать свои
``{\em Лекции по динамике}'' в Кёнигсберге. До этого, в 1838 году, Лиувилль просил одного из своих друзей перевести
для своего журнала (Journal de Math\'ematique pures et appliqu\'ees) две заметки Якоби по формализму Гамильтона--Якоби, опубликованных годом ранее в
журнале Крелля. В том виде в каком результаты Лиувилля известны сейчас, они были опубликованы только в середине 1850-х годов.
Между этими периодами была революция 1848 года и спад математической активности Лиувилля: он
был избран членом Национальной Ассамблеи (даже изложив  свои политические взгляды) и работал в комитете
по финансам.}. Тем не менее в начале 1830-х годов он
фактически первым начал классифицировать трансцендентные/алгебраические функции и четко формулировать понятие
<<проинтегрировать>> как конечный набор операций неопределенного интегрирования (знак интеграла $\int$) над вполне определенными
объектами (рациональные или алгебраические функции) + алгебраические действия над ними. Сейчас все это известно как
интегрирование в конечном виде (integration in finite terms) или
в замкнутой форме (closed form),
а применительно к линейным дифференциальным уравнениям --- как
теория Лиувилля--Пикара--Вессио и дифференциальная теория Галуа.
Например Лиувилль показал, что интеграл уравнения $\Psi''-x\,\Psi=0$
(уравнение Лиувилля--Эйри) не представим в элементарных функциях.
Последние разумеется были  определены.

В последующих параграфах мы покажем, что если речь идет о спектральной задаче (1), т.\,е. дифференциальном уравнении с параметром, то
обе интегрируемости совпадают, а все  перечисленные выше объекты появляются сами собой.
Таким образом естественная постановка вопроса выглядит следующим образом: 
\begin{itemize}
\item {\em При каких функциях $u(x)$ дифференциальное уравнение $(1)$ интегрируется в квадратурах при всех значениях параметра $\lambda$?\/} Класс функций
от которых эти интегралы вычисляются следует явно
описать.
\end{itemize}
Именно такая формулировка впервые встречается в работах 1919 года французского математика Ж.\,Драша (1871--1941). 

Еще раз отметим, что
терминология {\em точная интегрируемость\/} спектральных задач (линейных уравнений) и условий их совместности ($[L,A]$-пары и нелинейные уравнения на их коэффициенты) используется в литературе давно и в разных трактовках. Мы указываем на то, что все эти интегрируемости идентичны
<<наивному>> понятию, т.\,е. конечному набору значков интеграла,
и 
эквивалентны лиувиллевской интегрируемости в квадратурах
{\em линейных дифференциальных уравнений\/}, которая более элементарна чем  интегрируемость нелинейных гамильтоновых систем.

\section{Ж.\,Драш  и конечнозонные потенциалы}

\noindent
Даже беглый взгляд на формулы (см. приложение) показывает, что
им были решены принципиальные проблемы за исключением $\Theta$-функциональных
формул и гамильтоновости уравнений. Формулировка проблемы по Драшу совершенно прозрачна, элементарна и глубоко мотивирована. 

\subsection{<<Неспектральный>> взгляд на спектральные задачи}
\noindent
Объектом спектральных задач
является дифференциальное уравнение с параметром, которое надо решить (в каком-либо смысле) при всех значениях последнего.
Это диктуется требованием иметь полную систему собственных функций,  разложения по этому базису, ортогональность, разбиение единицы, равенства Парсеваля, спектральные матрицы-плотности и т.\,д\footnote{Спектральные задачи почти всегда возникают из потребностей разлагать произвольные функции (начальные условия и т.\,д.) по базису собственных функций  этих задач.
Это является полным аналогом разложений функций в ряды Фурье по
синусам и косинусам, как собственным функциям задачи
$\Psi''+\lambda^2\Psi=0$. Ортогональность влечет возможность
предъявить явные формулы для коэффициентов разложений.
Фактически такая постановка задачи для бесконечного интервала
оси $x\in (-\infty\ldots\infty)$ исследовалась в упомянутой работе Ахиезера (ДАН СССР (1961), {\bf 141}(2), 263--266).}. Одним словом,
в качестве ответа должны быть  более менее явные формулы для
функции $\Psi$ как функции двух переменных $x$ и $\lambda$. В противном случае мы будем иметь лишь частные примеры проинтегрированного уравнения (1) с конкретным коэффициентом $u(x)$ и конкретными $\lambda$. Таких примеров
к тому времени было известно не мало (например функции Эрмита), но вопрос о {\em произвольности\/} $\lambda$, как видим, является принципиальным. Драш требует разрешимость неопределенными квадратурами,  выделяя курсивом ``{\em \ldots\ оставляя параметр $h$ произвольным\/}''.

Разумеется можно привести большое количество примеров проинтегрированных уравнений с параметром\footnote{Например взяв наперед заданную функцию, содержащую $\lambda$, и построив линейное дифференциальное уравнение, которому она удовлетворяет: часто встречающаяся шутка по поводу трактовки метода {\em <<обратной>>\/} задачи.}. Однако если потребовать, чтобы способ вхождения $\lambda$  в уравнение был как в (1), то окажется что
это в высшей степени нетривиальная задача. Все такие немногочисленные примеры были известны наперечет и появление новых дало большой стимул развитию теории уравнения (1).
Именно {\sc мозр} и конечнозонное интегрирование дали такие новые примеры и, как мы увидим далее,  это {\em все возможные\/} случаи.
Это было объявлено Драшем открытым текстом, хотя и без доказательств. Что касается теории обратных спектральных задач, то этот класс потенциалов тоже дает ответ на все вопросы последней, если наложить <<разумные>> пожелания на аналитическую эффективность формул для матрицы рассеяния, коэффициентов
прохождения и т.п.

Далее, после такой <<неспектральной>> постановки задачи, Драш предъявил почти все
ключевые формулы описывающие ее решение:
1) весь класс конечнозонных потенциалов оператора Шредингера; 
2) алгебраические кривые для них; 3) уравнения Новикова и их интегралы; 4) уравнение на резольвенту и ее обобщение на трансценденты Пенлеве; 5) квадратурная формула для $\Psi(x;\lambda)$; 6) уравнения Дубровина на полюса $\Psi$-функции; 7) интегральная формула этих уравнений в виде задачи обращения Якоби; 8) формулы следов (потенциал как абелева функция); 9) результат действия преобразования Дарбу на алгебраическую кривую (добавление/удаление кратных точек ветвления). И это все в двух кратких заметках в Comptes Rendus.
Последующие работы Драша тоже посвящены квадратурной интегрируемости, но нелинейных уравнений. Больше к этой теме он не возвращался. К тому времени (1920-е годы) понятие квадратурной интегрируемости было довольно четко сформулировано, поэтому Драш
без пояснений использовал понятие группы рациональности уравнения
из теории Пикара--Вессио. Не имея возможности дополнительно комментировать эту обширную тему мы предложим другую мотивацию и методику
вывода квадратурной интегрируемости уравнения (1). Она даже более
известна чем упомянутая теория, более развита и алгоритмизирована. 
Это --- теория непрерывных симметрий Ли или, в современной терминологии, групповой анализ дифференциальных уравнений. К уравнению (1), насколько нам известно, этот подход не применялся. Его исторической мотивацией является тоже самое: как интегрировать дифференциальные уравнения в квадратурах?
В том виде в каком нам потребуется, сведения из этой теории
полностью содержатся в {\em ``Азбуке группового анализа''\/} Н.\,Ибрагимова.

В начале первой заметки Драш упоминает общую задачу Штур\-ма--Ли\-у\-вил\-ля
\begin{equation}\label{S}
\frac{d}{dz}\!\left( 
k(z)\frac{d\psi}{dz}\right)+\big(l(z)-\lambda\,g(z) \big)\,\psi=0
\end{equation}
и ее сводимость (тоже квадратурами) к уравнению (1). 
Иными словами, все что будет сделано для одного уравнения переписывается для другого.
Поскольку оба уравнения линейные, то связь между  переменными $\Psi$  и $\psi$ должна быть тоже линейной, а проблема сводится лишь к
замене независимой переменной $z\to x$. Это делается, 
как скорее всего имел в виду 
Драш, с помощью подстановки Лиувилля.
В самом деле, делая в уравнении \reff{S} замену переменной $z\to x$  и требуя чтобы уравнение приобрело канонический вид (1), мы получаем формулы перехода:
\begin{equation}\label{zx}
x=\int\limits^{\,\,z}\!\!\sqrt{\mfrac{g(z)}{k(z)}}\,dz,
\qquad
\Psi(x)=\sqrt[4]{g(z)k(z)}\,\psi(z),
\end{equation}
где $\Psi(x)$ удовлетворяет уравнению типа (1)
$$
\Psi_{\mathit{xx}}-
\left\{
\frac14\frac{k^2(z)}{g^2(z)}\left(\frac{g(z)}{k(z)} \right)_{\mathit{\!\!zz}}-
\frac{5}{16}\frac{k^3(z)}{g^3(z)}\left(\frac{g(z)}{k(z)}\right)_{\!\!z}+
\frac{k_{\mathit{zz}}(z)}{2g(z)}-\frac{{k_z}{}^{\!\!\!2}(z)}{4g(z)k(z)}-
\frac{l(z)}{g(z)}
\right\}\Psi=\lambda\,\Psi,
$$
в котором переменная $z$ должна быть заменена выражением $z=z(x)$, полученным обращением первой формулы в \reff{zx}.

\section{Квадратурная интегрируемость уравнения Шредингера}
\noindent
Для ответа на поставленный вопрос необходимо найти
группу/алгебру симметрий дифференциального уравнения (1).
Если она окажется разрешимой, то процедура интегрирования, т.\,е.
переход к новым <<интегрируемым>> переменным решается совершенно
стандартно, по известным методикам из теории непрерывных симметрий. Наше уравнение имеет порядок 2, поэтому достаточно
иметь 2-х параметрическую группу, т.\,к. такая всегда разрешима.
Поиск симметрий означает определить инфинитезимальный генератор группы $\widehat {\mathfrak{G}}$, т.\,е. найти функции $\zeta(x,\Psi),\;\eta(x,\Psi)$ его определяющие:
$$
\widehat {\boldsymbol{\mathfrak{G}}}=\zeta(x,\Psi)\,\partial_x+\eta(x,\Psi)\,\partial_\Psi.
$$
Выполняя такие действия (которые давно компьютеризированы), мы получаем
$$
\widehat {\boldsymbol{\mathfrak{G}}}=(a\,\Psi +R)\,\partial_x+\Big\{ a'\,\Psi^2+\Big(\mfrac12\,R'+\alpha
\Big)\,
\Psi+b\Big\}\,\partial_\Psi,
$$
где функции $a(x;\lambda),\,b(x;\lambda)$ есть любые решения уравнения (1), 
$\alpha$ --- любая константа,  а $R(x;\lambda)$ --- новая функция, удовлетворяющая уравнению 3-го порядка
\begin{equation}\label{hermit}
R'''-4\,(u+\lambda)\,R'-2\,u'\,R=0.
\end{equation}
Функции $a(x),\;b(x)$ ничего не дают, т.\,к. являются решениями
уравнения, которое подлежит интегрированию\footnote{Это естественно, т.\,к. линейное уравнение
всегда имеет симметрию сдвига $\Psi \to \Psi+\varepsilon\,\varphi(x)$ на произвольное решение  $\varphi(x)$ самого уравнения.}. Положим их равными
нулю. Оставшаяся константа $\alpha$ говорит о том, что
симметрии не исчезают и остается две одно-параметрические группы. Это в точности то что нам требуется:
$$
\widehat {\boldsymbol{\mathfrak{G}}}_1^{}=\Psi\,\partial_\Psi, \qquad
\widehat {\boldsymbol{\mathfrak{G}}}_2^{}=2\,R\,\partial_x+R'\,\Psi\,\partial_\Psi.
$$
Таким образом исходное уравнение можно считать проинтегрированным,
если считать проинтегрированным уравнение (\ref{hermit}).
На первый взгляд  уравнение (\ref{hermit}) сложнее исходного (1)
(см. комментарий в \S\,6.3).
Это фундаментальное уравнение конечно хорошо известно и многократно встречается
в современных работах по теории солитонов. В спектральных подходах оно известно как уравнение на ядро резольвенты, но мы будем избегать каких-либо дополнительных трактовок.
По видимому впервые оно возникло и обсуждалось в работе Лиувилля 1839 года \cite[стр.\,430--431]{liouville}
как уравнение на произведение двух решений
$R=\Psi_1\Psi_2$ причем в том же контексте --- интегрирование
линейного дифференциального уравнения 2-го порядка в явной форме.
Спектральный параметр у Лиувилля не присутствует и он
рассматривал интегрирование уравнения в контексте алгебраической интегрируемости: поиск случаев, когда решение будет алгебраической функцией от $x$. 

Вернемся к процедуре интегрирования. Точечная замена (способы поиска которой тоже отработаны \cite{eisenhart}) старых переменных $(x,\Psi)$ к новым 
<<интегрируемым>> $(z,w)$ имеет вид \cite[стр.\,111, случай $3^\circ$]{eisenhart} 
$$
z=\int\limits^{\,\,x}\!\!\frac{dx}{R(x;\lambda)}, \qquad
w=\ln\frac{\Psi}{\sqrt{R(x;\lambda)}}.
$$
В новых переменных, как это следует из теории Ли, уравнение (1) должно стать <<элементарно решаемым>>. После простых вычислений, получаем уравнение
\begin{equation}\label{order3}
w_{\mathit{zz}}+{w_z}^{\!\!\!2}=
-\mfrac12\,R\,R''+\mfrac14\,R'^2+(u+\lambda)\,R^2.
\end{equation}
Пока этого нельзя сказать но, как замечает Драш, уравнение (\ref{hermit}) имеет первый интеграл. Эта проинтегрированная форма имеет вид
\begin{equation}
\label{mu}
\mu^2=-\mfrac12\,R\,R''+\mfrac14\,R'^2+(u+\lambda)\,R^2,
\end{equation}
где $\mu^2$ --- константа интегрирования. Она совершенно произвольна и {\em ни от чего не зависит\/}, в том числе от $\lambda$ (!). 

Возможность проинтегрировать один раз всякое линейное уравнение хорошо известна. Это всегда делается через логарифмическую производную $\phi=R'/R$. Мы однако подчеркнем, что
при такой подстановке константа интегрирования не появляется в уравнении на новую переменную
$\phi$. Интеграл $\mu$  не связан с этой подстановкой, но
как отмечает Драш, такая интегрируемость связана с  самосопряженностью уравнения \reff{hermit}.
Используя  уравнение \reff{mu}, получаем что действительно, уравнение (\ref{order3})
легко интегрируется, поскольку принимает вид
$$
w_{\mathit{zz}}+{w_z}^{\!\!\!2}=\mu^2.
$$
Возвращаясь назад, к переменным $(x,\Psi)$, получаем формулу для $\Psi$:
\begin{equation}\label{Psi}
\Psi_\pm(x;\,\lambda)=\sqrt{R(x;\lambda)}\,
\exp\!\!\int\limits^{\,\,x}\!\!\frac{\pm\mu\,dx}{R(x;\lambda)}.
\end{equation}
Решение получено, но с точностью до решения уравнения \reff{mu}.
Это уравнение тоже хорошо известно. В частности, что его правая часть будет полиномом по $\lambda$, если таковым будет функция $R(x;\lambda)$. Такая полиномиальная формула для $R(x;\lambda)$, как <<хороший>> анзац, известна в современной литературе. Об этом же пишет и Драш. Забегая вперед (см. \S\,6.3) скажем, что зависимость $\Psi$-функции от $\lambda$ довольно сложна,
как это можно видеть из {\sc мозр} или общей теории конечнозонного интегрирования: она описывается на языке $\Theta$-функций и абелевых интегралов.

Тем не менее мы хотим показать, что и этот анзац методически выводим, если продолжать придерживаться идеологии интегрируемости в квадратурной форме.

\section{Как появляется фундаментальный полином по $\lambda$?} 
\subsection{Зависимость функции $R$ от $\lambda$}
\noindent
Мы рассматриваем уравнение (1) как алгебраическое (полиномиальное) по $\lambda$  и дифференциальное по $x$. 
Не обсуждая отдельно, будем считать все зависимости функций $R$ и $\Psi$ от переменных $(\lambda,x)$ аналитическими. Значит как $\Psi$ так и $R$ разложимы в ряды Лорана--Тейлора по переменной $\lambda$. Поскольку мы логически пришли от $\Psi$-функции к функции $R$, напишем для нее, как и для любой аналитической функции вообще, ряд Лорана. В качестве точки $\lambda_0^{}$, в окрестности которой мы намереваемся разлагать функции, можно взять любую конечную точку плоскости $(\lambda)$. Но в таких случаях все дальнейшие формулы будут содержать эту <<нормировочную>> (или <<начальную>>) точку $\lambda_0^{}$. Чтобы избежать этого, мы берем точку $\lambda_0^{}=\infty$, получив при этом  разложения, справедливые всюду в конечной части плоскости $(\lambda)$. Добавим при этом, что будучи полиномиальным по $\lambda$, наше уравнение (1) имеет единственную особенность на бесконечности. Иными словами, это объясняет выделенную роль точки $\lambda=\infty$, а разложения следует брать  именно в ее окрестности\footnote{Если бы спектральная задача содержала бы особенности в некоторых конечных точках $\lambda_k$, выше приведенные рассуждения следовало бы видоизменить и рассматривать разложения в окрестности этих особенностей. В современных теориях такие случаи встречаются и соответствуют теориям с так называемыми многоточечными функциями Бейкера--Ахиезера.}.
Впервые такие разложения стали  делать Гельфанд и Дикий (1975), но мы отметим что эти ряды следует рассматривать не как формальные, а как реальные аналитические ряды:
$$
R(x;\lambda)=R_0(x)+R_1(x)\,\zeta+
R_2(x)\,\zeta^2+\cdots
+ R_k(x)\,\zeta^k+\cdots\;, \quad\mbox{где $\zeta=\displaystyle\frac{1}{\lambda}$}.
$$
Далее в теории хорошо появляются хорошо известные рекуррентные соотношения на коэффициенты $R_k$, интегральный оператор рекурсии, дифференциальные полиномы по $u(x)$ и т.\,п.
Мы не будем обсуждать эти детали, но продолжим следовать основной мотивации. Без потери общности (!), общий вид 
зависимости функции $R$ от обеих переменных получается следующим
$$
R(x;\lambda)=1-
\Big(\mfrac{1}{2}^{}\,u-c_1^{}\Big)\zeta-
\Big(\mfrac {1}{8}\,u_{\mathit{xx}}-\mfrac38\,u^2+
\mfrac12\,c_1^{}u-c_2^{}\Big)\zeta^2+\cdots\;.
$$
Новые константы $c_k$ появились, так как рекурренции содержат интегрирование по $x$ на каждом шаге. Отметим пока, что считая потенциал $u(x)$ зафиксированным, вся зависимость
от всех переменных функции $R$, включая зависимость от трех констант интегрирования $C_{1,2,3}$ уравнения (\ref{hermit}), которому она удовлетворяет, содержится в этой формуле и в  константах $c_k$. Величина $\mu$ тоже зависит от $C_{1,2,3}$, либо ее можно считать, в зависимости от выбора $C_{1,2,3}$, одной из этих констант интегрирования.
Все эти зависимости пока сложны и неизвестны, а потенциал $u(x)$
совершенно произволен и ограничен лишь требованием быть аналитической функцией. Аналитичность всех функций мы  будем подразумевать всюду и в дальнейших рассуждениях. Подставив последнее разложение
в уравнение (\ref{mu}) мы получим
$$
\zeta\,\mu^2=1+2\,c_1^{}\zeta+\big(2\,c_2^{}+c_1^2\big)\zeta^2+
2\,\big(c_3^{}+c_1^{} c_2^{}\big)\zeta^3_{\mathstrut}+
\cdots+\mbox{оставшаяся часть},
$$
где <<оставшаяся часть>> означает те слагаемые, которые содержат потенциал $u(x)$ и его производные $u^{(k)}$.
Ключевым моментом является то, что до тех пор пока функция $R$
определяется бесконечным рядом по $\lambda$, т.\,е. трансцендентна, первая часть, содержащая только константы $c_k$, в последнем уравнении будет бесконечна. Бесконечным будет и порядок производных от $u$ во второй части. Весь этот ряд, по построению,  есть константа $\zeta\mu^2$, а требование быть константой при любых $\lambda$ говорит о том что константой будет каждый коэффициент перед $\zeta$. Одним словом вся зависимость от $u(x)$ уходит в бесконечность. В то же время вполне понятно, что квадратурная интегрируемость имеет место не при любых потенциалах.
Чтобы получить условия на потенциал конечного порядка (и даже получить хоть какие-то условия) мы приходим к единственной необходимости:  положить ряд конечным. 
\begin{itemize}
\item {\em Фундаментальная функция $R(x;\lambda)$ должна быть полиномом по $\lambda$\/}:
\begin{equation}\label{poly}
R(\pot;\lambda)=\lambda^g+R_1(\pot)\lambda^{g-1}+\cdots+R_{g-1}(\pot) \lambda+R_g(\pot).
\end{equation}
\end{itemize}
Начиная с этого момента мы будем отмечать зависимость функции
$R$ от $x$ как $R(\pot;\lambda)$ имея ввиду то что она становится дифференциальным полиномом от потенциала.
Таким образом полиномиальность этой функции есть не просто простейшая зависимость, но и единственная возможность.
Для справок приведем
рекуррентные формулы для коэффициентов $R_k$:
$$
\mbox{\small$\displaystyle
\begin{array}{c}
\displaystyle R(\pot;\lambda)=\sum_{n=0}^g \lambda^n  \sum_{j=0}^{g-n}\, \,c_j\,R_{g-n-j},
\quad c_0^{}=R_0^{}=1,\quad R_1^{}=-\mfrac{1}{2}\,u,\\
\displaystyle \!\!\!\!\! R_k=\mfrac{1}{8}\,\sum_{j=1}^{k-1} \left(
2\,R''_j\,R_{k-j-1} -R'_j\,R'_{k-j-1} - 4\,R_j\,R_{k-j} - 4\,u\,R_j\,R_{k-j-1}
\right)- \mfrac12\,u\,R_{k-1}\quad (k>1).
\end{array}
$}
$$
\subsection{Появление алгебраической кривой}
\noindent
Однократно проинтегрированная форма \reff{mu} дифференциального уравнения \reff{hermit} становится выражением вида
\begin{equation}\label{curve}
\mu^2=\lambda^{2g+1}+I_1(c_k^{})\,\lambda^{2g}+\cdots
+I_{2g}(\pot;c_k^{})\,\lambda+I_{2g+1}(\pot;c_k^{}).
\end{equation}
Если по прежнему считать $\mu$ величиной не зависящей от $\lambda$, тогда выражение \reff{curve} будет представлять собой дифференциальное уравнение на потенциал, содержащее $\lambda$.
Это не допустимо, так как потенциал не должен зависеть от $\lambda$. Значит величина $\mu$ должна стать зависимой от $\lambda$ посредством этого уравнения, а само $\lambda$ по прежнему остается произвольным.
Следовательно каждый коэффициент $I_k$ должен быть константой $A_k$ не зависящей от $\lambda$ и, таким образом, порождать
дифференциальное уравнение $I_k(\pot;c)=A_k$ конечного порядка 
на потенциал.
Эти уравнения, которые называются уравнениями Новикова, не противоречат друг другу, т.\,к. мы знаем что интегрируемые потенциалы существуют (например, все солитоны).
Набор коэффициентов $I_k$ распадается на два множества:
константы $c_k^{}$ и собственно коэффициенты выражающиеся через
потенциал $u(x)$ и его производные конечного порядка.

Что касается $\mu$ как константы интегрирования, то ее зависимость от $\lambda$ указанного вида означает
выбор {\em частного\/} и {\em полиномиального\/} решения уравнения \reff{hermit}. Она зависит еще и от потенциала через константы $A_k$ и $c_k^{}$. Уравнение \reff{curve} представляет собой первый фундаментальный объект теории конечнозонного интегрирования: алгебраическую кривую $W(\lambda,\mu)=0$, связывающую спектральный параметр $\lambda$  и параметр $\mu$.  В современных формулировках эта кривая, известная как спектральная кривая, часто выступает как первичный объект, а дальнейшая теория строится по ней аксиоматически.

\subsection{Соотношение между $\Psi$-функцией и функцией $R$}
\noindent
Возможно ли вывести теорию, включая ее дальнейшее
изложение, используя только само уравнение (1) не прибегая к функции $R$? По всей видимости нет. Из уравнения (1) мы имеем, что логарифмическая производная $\phi=\Psi_x/\Psi$ удовлетворяет уравнению Риккати
\begin{equation}\label{riccati1}
\phi_x(x;\lambda)+\phi(x;\lambda)^2=u(x)+\lambda,
\end{equation}
а выражение $\phi_x+\phi^2$ есть линейная функция от $\lambda$.
Какой в этом случае  должна быть зависимость $\phi$ от $\lambda$?
Простые догадки не приводят к успеху, однако из формулы \reff{Psi}
\begin{equation}\label{riccati2}
\phi(x;\lambda)=\frac{\mu(\lambda)+\frac12 R'(\pot;\lambda)}{R(\pot;\lambda)}
\end{equation}
мы видим что $\phi(x;\lambda)$ есть довольно специального вида
алгебраическая функция от $\lambda$ принадлежащая гиперэллиптической иррациональности \reff{curve}: отношение
числителя $\mu(\lambda)+\frac12R_x(\lambda)$ и знаменателя $R(\lambda)$. Коэффициентами в них являются сложные зависимости от потенциала, а сам потенциал удовлетворяет
нелинейным уравнениям, содержащим дополнительные константы $c_k$.
Иными словами, в терминах $R$-функции, $\lambda$-зависимость проста, а появление самой этой функции является неизбежным атрибутом теории.

\subsection{Примеры}
\noindent
В качестве демонстрации приведем полную формулу когда порядок полинома
$R$ равен единице ($R(\pot;\lambda)=\lambda-\frac12\,u+c_1^{}$):
$$
\mu^2=\lambda^3+2\,c_1^{}\lambda^2+\mfrac14\, 
\big(u_{\mathit xx}-3\,u^2+4\,c_1^{}u+4\,c_1^2\big) \,\lambda-
\mfrac18\,(u-2\,c_1^{})\,u_{\mathit{xx}}
+\mfrac{1}{16}\,u_x^2+\mfrac14\,u\,(u-2\,c_1^{})^2.
$$
При $c_1^{}=0$ эта формула превращается в выражение \reff{g1}.
Такой выбор константы всегда допустим.  Величина $\lambda$ произвольна и ее можно заменить на $\lambda-\frac23\, c_1^{}$. Если также сдвинуть потенциал $u\to u+\frac23\,c_1^{}$, то получим в точности формулу \reff{g1}. 

Соотношения $I_k(\pot;c)=A_k$ не просто представляют собой дифференциальные уравнения, но уравнения уже на половину проинтегрированные с константами интегрирования $A_k$. Наблюдение Новикова состояло в том, что
это как раз та половина интегралов, которая необходима в лиувиллевской интегрируемости нелинейной гамильтоновой системы. Рассмотренный нами пример $g=1$ легко интегрируется и без всякой теории. Полагая $c_1^{}=0\;\to I_1=0$, получаем
$$
I_2(\pot)=A_2:\quad u_{\mathit xx}-3\,u^2=4A_2,\qquad\quad
I_3(\pot)=A_3:\quad -\mfrac18\,u\,u_{\mathit{xx}}
+\mfrac{1}{16}\,u_x^2+\mfrac14\,u^3=A_3.
$$
Подставляя $u_{\mathit{xx}}$ из одного интеграла в другой и интегрируя, приходим к функции Вейерштрасса $u(x)=2\wp(x+a;g_2^{},g_3^{})$, где константы $A_{2,3}$ заменены новыми константами $g_{2,3}^{}$. Это хорошо известная кноидальная волна Кортевега--де Фриза, частным случаем которой является солитон. Эта же пара уравнений на потенциал, после дифференцирования любого уравнения из нее, т.\,е. исключения констант $A_{2,3}$, эквивалентна одному уравнению $u_{\mathit{xxx}}-6\,u\,u_x=0$, которое совпадает со стационарной версией уравнения КдФ (2).

С ростом степени полинома $R$  формулы быстро разрастаются, но весь процесс легко алгоритмизируется. 
Для степени $g=2$ получаем выражения типа следующих
\begin{equation}\label{g2}
\mbox{\small$
\begin{array}{l}
\mu^2=\lambda^5+2\,c_1^{}\,\lambda^4+
(2\,c_2^{}+c_1^2)\,\lambda^3+\\
\phantom{\mu^2}
+\mfrac{1^{\mathstrut}}{16}\,
\big(u_{\mathit{xxxx}}-10\,u\,u_{\mathit{xx}}-5\,u_x^2
+10\,u^3+4\,c_1^{}(u_\mathit{xx}-3\,u^2)+16\,c_2^{}(u+2\,c_1^{})
\big)
\,\lambda^2-\cdots
\end{array}
$}.
\end{equation}
Уравнение Новикова, которому удовлетворяет потенциал, имеет здесь вид
$$
u_{\mathit{xxxxx}}-10\,u\,u_{\mathit{xxx}}-20\,u_x u_{\mathit{xx}}+30\,u^2u_x+4\,c_1^{}(u_{\mathit{xxx}}-6\,u\,u_x)
+16\,c_2^{}u_x=0.
$$
В общем случае ситуация выглядит аналогично. Фиксируя степень полинома равной $g$, получаем стационарное уравнение Новикова порядка  $2g+1$ с дополнительными константами $c_k^{}$ (но без констант $A_k$).  Уравнение кривой \reff{curve} содержит $g$ констант $c_k^{}$ и $g+1$
констант $A_k$ --- интегралов уравнения Новикова.
Это уравнение может быть переписано в гамильтоновой форме как динамическая система по переменной $x$ с размерностью фазового пространства равной $2g$. Она интегрируема по Лиувиллю.

\subsection{Эквивалентность <<линейной>> и <<нелинейной>>
интегрируемости}
\noindent
Главный вывод, который мы сейчас имеем, состоит в том что требование квадратурной интегрируемости {\em линейного\/} уравнения (1)  свелось к условиям на потенциал в виде квадратурной интегрируемости {\em нелинейных\/} дифференциальных уравнений Новикова. Последние же являются гамильтоновой системой (см. формулы в \cite{novikov}).

Важно отметить, что на данный момент проблема  еще не решена, даже при наличии формулы \reff{Psi}  для $\Psi$-функции. Ее следует рассматривать как {\em анзац\/} и таковым она будет оставаться до тех пор пока не будет предъявлена
квадратурная формула для потенциала $u(x)$, входящего в
функцию $R$. Решение уравнения (1) при {\em любом\/} потенциале будет иметь вид \reff{Psi}. Половину интегралов мы уже имеем. Возможно ли извлечь оставшуюся и, если да, то, как выглядит ответ?
Забегая вперед скажем что нам предстоит прийти к задаче обращения
Якоби, знаменитым тождествам следов и  переменным $\gamma_k^{}$. $\Theta$-функции последуют как представления для решений.

\section{Как появляются корни фундаментального полинома $R=\big(\lambda-\gamma_1^{}(x)\big)\cdots
\big(\lambda-\gamma_g(x)\big)$?}

\noindent
Несмотря на то что наши уравнения наполовину проинтегрированы
и написана формула для $\Psi$-функции мы вернемся к первичным объектам \reff{riccati1} и \reff{riccati2}. Они не связаны
ни со свойствами потенциала быть интегрируемым, т.\,е. конечнозонным, ни с полиномиальностью функции $R(\pot;\lambda)$. 
Уравнение Риккати \reff{riccati1}  с учетом появившейся фундаментальной функции $R$ имеет вид (подставляя \reff{riccati2} в \reff{riccati1})
\begin{equation}\label{polar}
\frac{\mu^2}{R^2}+\frac12\,\frac{R''}{R}-\frac14\,\frac{R'^2}{R^2}=
u(x)+\lambda.
\end{equation}
Данное уравнение (т.\,е. уравнение \reff{mu}) еще не проинтегрировано.  По всей видимости Драш руководствовался рассуждениями, которые излагаются ниже (см.  перевод его второй статьи).
Отметим еще раз, что величина $\mu$, в данный момент, как константа интегрирования не зависима, в первую очередь от $\lambda$. 

Рассмотрим  формулу  \reff{polar} как функцию от $\lambda$. Ее правая часть есть линейная функция от $\lambda$. Таковой же должна быть и левая часть, которая однако может обращаться в бесконечность  в точках $\lambda_k$, где $R$ обращается в нуль. 
Так как $\lambda$ произвольно (!), мы можем устремить $\lambda$ к этим точкам не зависимо от того, что сами эти точки $\lambda_k=\gamma_k(x)$ являются функциями переменной
$x$, которая выступает здесь как параметр.
Количество этих нулей функции $R$ тоже не имеет значения. Их может быть конечное или бесконечное число. Все зависит от того 
насколько <<хорошим>> выбран потенциал.
Таким образом нам следует написать $\lambda$-разложения  выражения \reff{polar} в окрестности каждого нуля $\gamma_k(x)$ и потребовать обращения
каждого коэффициента  в ноль.
Важно отметить, что данное рассуждение совершенно независимо 
от предыдущего параграфа и может предшествовать ему.
Там мы пришли к полиномиальности функции $R$ и уравнениям Новикова на потенциал. Здесь же, разложения
для каждого нуля приведут к самостоятельным условиям на
$x$-зависимость функции $R(\pot;\lambda)$ посредством новых величин $\gamma_k(x)$. Так как уравнение \reff{polar} есть дифференциальное уравнение на $R$, то этими условиями должны стать дифференциальные уравнения на переменные $\gamma_k^{}$.  Если нулей конечное число, то конечным будет и количество таких уравнений. Драш заметил, что этих уравнений
будет $g$ штук, все они первого порядка, автономны и явно интегрируются.
Таким образом фундаментальность $R$-функции двояка: полиномиальность по $\lambda$ влечет {\em необходимые\/} условия интегрируемости, а нули $\gamma_k^{}$ реализуют {\em достаточность\/}, т.\,е. процедуру интегрирования. 

\section{Процедура интегрирования}
\noindent
Суммируя предыдущее мы имеем полиномиальную функцию 
$$
R(\pot;\lambda)=\big(\lambda-\gamma_1^{}(x)\big)\cdots
\big(\lambda-\gamma_g(x)\big),
$$
после чего уравнение кривой \reff{mu}, будучи дифференциальным уравнением 2-го порядка на функцию $R$,  в силу произвольности $\lambda$ (!),
{\em расщепляется\/} на $g+1$ уравнений на потенциал \reff{curve},
давая при этом интегралы уравнения Новикова. Эти уравнения подлежат дальнейшему интегрированию. Чтобы избежать излишней общности мы опишем
дальнейшие действия на нетривиальном примере, когда $g=2$.
Обобщения на произвольные $g$ не составляют никакого труда.

Заметим только что случай $g=1$ в действительности не показателен и элементарен
в том смысле что вообще не требует никаких теорий, так как нет необходимости вводить переменную $\gamma$: как мы видели ранее, уравнение Новикова в этом случае интегрируется сразу.
Вероятно по этой причине этот случай долгое время оставался единственным
явно проинтегрированным примером.

\subsection{Интегрирование случая $g=2$}
\noindent
Напишем формулу
$$
\begin{array}{l}
R(\pot;\lambda)=\lambda^2+\Big(c_1^{}-\mfrac12\, u \Big)\,\lambda-
\mfrac18\,u_{\mathit{xx}}+\mfrac38\,u^2-\mfrac12\,c_1^{}u+c_2^{}=\\
\phantom{R(\pot;\lambda)}=(\lambda-\gamma_1(x))(\lambda-\gamma_2(x))
^{\displaystyle{}^{\mathstrut}}_{}=
\lambda^2-(\gamma_1^{}(x)+\gamma_2^{}(x))\,\lambda+
\gamma_1^{}(x)\gamma_2^{}(x)
\end{array}
$$
из которой следует связь между потенциалом  и появившимися переменными $\gamma_{1,2}^{}(x)$:
\begin{equation}\label{trace}
c_1-\frac12\,u=-\gamma_1^{}(x)-\gamma_2^{}(x).
\end{equation}
Это знаменитая формула следов. Если будут найдены величины $\gamma_{1,2}^{}$, то потенциал находится по этой формуле.
Упомянутые условия на $x$-зависимость функции $R$ через величины
$\gamma_{1,2}^{}(x)$ {\em извлекаются из  уравнения кривой\/}, т.\,е. формулы \reff{curve} или, что удобней, из \reff{polar}. Поскольку мы находимся в конечнозонном случае, величина $\mu$ зависит от $\lambda$ и в пределе $\lambda\to\gamma_k^{}(x)$ станет алгебраической функцией
$\mu\to \nu_k^{}(x)$,  определяемой уравнением кривой \reff{g2}:
$$
\begin{array}{l}
\nu_k^2(x)=\gamma_k^5(x)+2\,c_1^{}\gamma_k^4(x)+
(2\,c_2^{}+c_1^2)\,\gamma_k^3(x)+A_3\gamma_k^2(x)+
A_4\gamma_k^{}(x)+A_5=\\
\phantom{\nu_k^2(x)}=(\gamma_k^{}(x)-E_1)\cdots(\gamma_k^{}(x)-E_5)^
{\displaystyle{}^{\mathstrut}}_{}.
\end{array}
$$
Разлагая уравнение \reff{polar} в окрестности точек $\gamma_{1,2}^{}(x)$ получаем
$$
\mbox{\small$\displaystyle
\left\{\frac{\nu_1^2}{(\gamma_1^{}-\gamma_2^{})^2}
-\frac14\,\gamma_1'^2\right\}
\frac{1}{(\lambda-\gamma_1^{})^2}-
\left\{
\frac{\nu_1^2}{(\gamma_1-\gamma_2^{})^3}+\frac12\,\gamma_1''
-\frac12\,\frac{\gamma_1'\gamma_2'}{\gamma_1-\gamma_2}\right\}
\frac{1}{(\lambda-\gamma_1^{})}+\cdots=0
$},
$$
$$
\mbox{\small$\displaystyle
\left\{\frac{\nu_2^2}{(\gamma_2^{}-\gamma_1^{})^2}
-\frac14\,\gamma_2'^2\right\}
\frac{1}{(\lambda-\gamma_2^{})^2}-
\left\{
\frac{\nu_2^2}{(\gamma_2-\gamma_1^{})^3}+\frac12\,\gamma_2''
-\frac12\,\frac{\gamma_2'\gamma_1'}{\gamma_2-\gamma_1}\right\}
\frac{1}{(\lambda-\gamma_2^{})}+\cdots=0
$},
$$
где многоточия означают высшие члены разложений, содержащие
$\gamma_{1,2}^{},\gamma_{1,2}', \gamma_{1,2}''$ и сам потенциал $u(x)$. Старшие члены разложения дают
\begin{equation}\label{dubrovin}
\frac{d\gamma_1^{}}{dx}=2\,\frac{\nu_1^{}}{\gamma_1^{}-\gamma_2^{}},
\qquad
\frac{d\gamma_2^{}}{dx}=2\,\frac{\nu_2^{}}{\gamma_2^{}-\gamma_1^{}}.
\end{equation}
Это дифференциальные автономные уравнения первого порядка  уравнения, известные как уравнения Дубровина. 
Важным результатом Драшa является то что они легко интегрируются. В самом деле, записывая их в виде
$$
\frac{d\gamma_1^{}}{\nu_1^{}}=\frac{2\,dx}{\gamma_1^{}-\gamma_2^{}},
\qquad
\frac{d\gamma_2^{}}{\nu_2^{}}=\frac{2\,dx}{\gamma_2^{}-\gamma_1^{}},
$$
получаем, что
$$
\frac{d\gamma_1^{}}{\nu_1^{}}+\frac{d\gamma_2^{}}{\nu_2^{}}=0
\qquad\mbox{и}\qquad
\frac{\gamma_1^{}d\gamma_1^{}}{\nu_1^{}}+
\frac{\gamma_2^{}d\gamma_2^{}}{\nu_2^{}}=2\,dx.
$$
Хотя еще не ясно как выглядят функции $\gamma_{1,2}^{}(x)$ 
мы получили главное --- эти уравнения уже выглядят как {\em интегрируемые в неопределенных квадратурах\/}:
\begin{equation}\label{jacobi}
\int\limits^{\,\gamma_1^{}}\!\frac{d\lambda}{\mu}+
\int\limits^{\,\gamma_2^{}}\!\frac{d\lambda}{\mu}=a_1^{},
\qquad
\int\limits^{\,\gamma_1^{}}\!\frac{\lambda d\lambda}{\mu}+
\int\limits^{\,\gamma_2^{}}\!\frac{\lambda d\lambda}{\mu}=
2\,x+a_2^{}.
\end{equation}
Процедура интегрирования завершена: появились две недостающие константы интегрирования $a_{1,2}^{}$, а сами величины $\gamma_{1,2}^{}$,  как функции от $x$, находятся путем {\em обращения\/} этих  интегралов: верхние пределы $\gamma_{1,2}^{}$ следует определить как функции от значений двух сумм интегралов $a_1^{}$ и $2\,x+a_2^{}$.
В уравнениях \reff{jacobi} величина $\mu=\mu(\lambda)$ определяется формулой 
\begin{equation}\label{curve2}
\mu=\pm\sqrt{(\lambda-E_1)\cdots(\lambda-E_5)}
\end{equation}
и, таким образом, все интегралы являются абелевыми интегралами принадлежащими этой гиперэллиптической иррациональности.  Нетрудно догадаться что
дальнейшие члены разложения уравнения \reff{polar} удовлетворяются автоматически. В действительности они определяют полную совокупность соотношений между переменными $u(x),\gamma_k^{}(x), \gamma_k'(x),\gamma_k''(x)$.  
В этот набор войдут как уравнения
Драша--Дубровина \reff{dubrovin} так и  формула следов \reff{trace}, независимо от того что она уже извлеклась из полинома $R$.
Формулы (\ref{trace}--\ref{dubrovin}) представляют, помимо уравнения кривой
\reff{curve2}, другие два фундаментальных объекта теории
конечнозонного интегрирования.

\subsection{Общее решение задачи}
\noindent
Подитожим результаты выписывая формулы для произвольного $g$. Интегрирование в неопределенных
квадратурах спектральной задачи, определяемой уравнением Шредингера (1), возможно только в случаях когда функция-потенциал
$u(x)$ удовлетворяет дифференциальным уравнениям Новикова $I(\pot;c)=A_k$, фигурирующим в уравнении алгебраической кривой \reff{curve}. Решение-анзац для  $\Psi$-функции дается выражением \reff{Psi}, в котором функция $R(\pot;\lambda)$
является полиномом \reff{poly}. Вместо потенциала $u(x)$
в данную функцию и формулу \reff{Psi} следует подставить
формулу следов
$$
u=2\sum\limits_{k=1}^{g}\gamma_k^{}(x)-\sum\limits_{k=1}^{2g+1}E_k,
$$
в которой функции $\gamma_k^{}=\gamma_k^{}(x)$ определяются
из решения задачи обращения Якоби
\begin{equation}\label{Jacobi}
\begin{array}{cl}
\displaystyle
\int\limits^{\,\gamma_1^{}}\!\frac{d\lambda}{\mu}+\cdots+
\int\limits^{\,\gamma_g^{}}\!\frac{d\lambda}{\mu}&=a_1^{}\\
\displaystyle
\int\limits^{\,\gamma_1^{\mathstrut}}\!\frac{\lambda d\lambda}{\mu}+\cdots+
\int\limits^{\,\gamma_g^{}}\!\frac{\lambda d\lambda}{\mu}&=a_2^{}\\
\cdots\cdots\cdots&\cdots^{\displaystyle\mathstrut}_{}\\
\displaystyle
\int\limits^{\,\gamma_1^{}}\!\frac{\lambda^{g-1} d\lambda}{\mu}+\cdots+
\int\limits^{\,\gamma_g^{}}\!\frac{\lambda^{g-1} d\lambda}{\mu}&=
a_g+2\,\;.
\end{array}
\end{equation}
для гиперэллиптической алгебраической кривой \begin{equation}\label{hyper}
\mu^2=(\lambda-E_1)\cdots(\lambda-E_{2g+1})
\end{equation}
Окончательная $\Psi$-формула в которую превращается упомянутый анзац \reff{Psi} имеет вид
\begin{equation}\label{final}
\Psi_\pm(x;\lambda)=C\cdot\exp\frac12
\left\{
\int\limits^{\gamma_1^{}(x)}\!\frac{w \pm\mu}{z-\lambda}\,
\frac{dz}{w}+
\cdots+
\int\limits^{\gamma_g^{}(x)}\!\frac{w \pm\mu}{z-\lambda}\,
\frac{dz}{w}
\right\},
\end{equation}
где $w=w(z)$ связано с переменной интегрирования соотношением 
$w^2=(z-E_1)\cdots(z-E_{2g+1})$.
Эта формула по всей видимости отсутствует в литературе, но почти появилась у Драша.
Остается заметить, что она может быть проверена прямой
подстановкой в уравнение (1) и использованием формулы следов и уравнений Драша--Дубровина
$$
\frac{d\gamma_k^{}}{dx}=2\,\frac{\sqrt{(\gamma_k^{}-E_1)\cdots
(\gamma_k^{}-E_{2g+1})}}{\displaystyle\prod\limits_{j\ne k}^g(\gamma_k^{}-\gamma_j^{})},
$$
которые являются просто дифференциальной формой интегральной задачи обращения \reff{Jacobi}.
Сам потенциал определяется дополнительными константами $c_{1\ldots g}^{}$, входящими в уравнения Новикова. Последние, как уравнения порядка $2g+1$, имеют в качестве интегралов
$g+1$ коэффициентов кривой $A_k$ и $g$ величин $a_{1\ldots g}^{}$, фигурирующих в задаче Якоби \reff{Jacobi}.

Добавим еще что  нет необходимости
в отдельных формулах для верхних пределов $\gamma_k^{}(x)$  абелевых интегралов \reff{Jacobi}.
Для решения задачи достаточно знать лишь сумму этих величин.
Подобные симметрические выражения, как и всевозможные 
рациональные симметрические комбинации
величин $\gamma_k^{}$ и $\nu_k^{}$, как функции от значений интегралов, называются абелевыми функциями
и выражаются через $\Theta$-функции Римана. Сама $\Psi$-функция,
являясь симметрической по переменным $\gamma_k^{}(x)$,
не является абелевой функцией\footnote{Функция называется абелевой если она симметрична по переменным $(\gamma_k^{},\nu_k^{})$
и является мероморфной функцией от значений интегралов, фигурирующих в задаче Якоби \reff{Jacobi}. Наша $\Psi$-функция
имеет экспоненциальную зависимость и потому не является абелевой.}, но тоже допускает представление  в терминах $\Theta$-функций. Такая $\Psi$-функция, в которой $\lambda$-зависимость фигурирует  как первичная (спектральный подход), а $x$ --- параметр, называется функцией Бейкера--Ахиезера.

Для случая $g=1$, в принятой выше нормировке $c_1^{}=0$, мы получаем ответ
$$
u=2\,\gamma(x), \quad \mbox{где} \quad
\int\limits^{\gamma}\!\!
\frac{d\lambda}{\sqrt{\lambda^3+A_2\lambda+A_3}}=2\,x+2\,a,
$$
который эквивалентен   формулам с функцией Вейерштрасса: $\gamma=\wp(x+a;-4A_2,-4A_3)$. 

\section{Процедура обращения}

\noindent
Вообще говоря, нетривиальный момент, связанный с обращением абелевых интегралов, имеет лишь {\em кажущуюся\/} сложность.
Он неизбежно присутствует даже в процедуре интегрирования тривиального потенциала $u(x)=\mathit{const}$. При этом совершенно не важно какими средствами осуществляется интегрирование поскольку нет выделенной роли зависимых/не\-за\-ви\-си\-мых переменных. В самом деле, проинтегрируем уравнение
$$
\Psi''=A\,\Psi,
$$
соответствующее постоянному потенциалу, элементарными средствами без привлечения каких-либо теорий, но не вводя никаких <<новых>> функций типа $\sin$ или $\arcsin$. Домножая это уравнение на
$\Psi'$ и интегрируя, получаем
$\Psi'^2=A\,\Psi^2+C_1$.
Разделяя переменные имеем
$$
\int\!\!\frac{d\Psi}{\sqrt{A\Psi^2+C_1}}=\int \!\! dx +C_2.
$$
Уравнение проинтегрировано.
Привлекая {\em рациональную\/} замену переменной $\Psi\to z$  типа $\Psi=c\,(z+z^{-1})$ (параметризация окружности),
после алгебраических действий приходим к необходимости
вычислять и {\em обращать\/} интеграл: 
$$
\int\limits^{\,z}\!\frac{dz}{z}=\int\limits^{\,\,x} \!a\,dx \quad\to\quad \ln z=ax+b\quad
\ldots\quad\to \quad \Psi_\pm= e^{\pm\sqrt{\!A}\,x}.
$$
Иными словами, в этом предельно вырожденном случае абелевы интегралы не возникают, а точнее, возникают {\em тривиальные\/} абелевы интегралы --- интегралы от рациональных функций, которые всегда берутся в логарифмах. 
Как  собственно в конечнозонном случае, так и в рассматриваемом
примере, уравнения интегрируются в стандартных квадратурах только после перехода к новым переменным. 
Сам логарифм  следует рассматривать как {\em определяемый\/} через
интеграл от простейшей рациональной функции
$$
\ln z\stackrel{\mathrm{Def}}{\equiv}\int\limits^{\,z}\!
\frac{dz}{z},
$$
а экспоненту $\exp(x)$ --- как целую трансцендентную периодическую функцию  одной переменной {\em определяемую\/}
его обращением. 
Ее периодичность следует из свойства интеграла получать приращение $2\pi i$ после обхода особой точки интеграла.
Это единственный возникающий период.

В более сложных случаях мы приходим к необходимости обращать собственно абелевы интегралы. Поскольку решение свелось
к конечному числу интегрирований алгебраических функций, в другой терминологии теория известна как {\em алгебраическая интегрируемость\/}.
Все оставшиеся  действия имеют отношение лишь к 
дальнейшему {\em представлению\/}  интегралов и их обращений, 
но не затрагивают природу интегрируемости. Сами представления реализуются с помощью новых целых трансцендентных функций, но уже от нескольких переменных, т.\,к. интегралов несколько.
Этими функциями являются знаменитые $\Theta$-ряды\footnote{Тот факт что $\Theta$-функции лежат в основе конечнозонного интегрирования является по всей видимости общепринятой точкой зрения.
С другой стороны, аксиоматически постулируемый $\Theta$-ряд не может объяснять какую-либо интегрируемость.}.
Их свойства периодичности тоже следуют из свойств абелевых интегралов \reff{Jacobi} получать приращения. Однако число контуров, самих интегралов,
и их независимых периодов будет уже больше. Все это отразится на 
свойствах периодичности трансцендентного $\Theta$-ряда
как функции от $g$ переменных. Иными словами, неизбежность появления новых экспоненциальных трансцендент
такая же как и появление экспоненты в вырожденном случае.
Как результат, новые трансценденты должны быть {\em введены\/}, а выражения для самих рядов строятся не аксиоматически постулируемой, а регулярной процедурой. Объяснение последнего важного замечания выходит за рамки данной заметки.

Похожие явления, как известно, имеют место и при интегрирования
по Лиувиллю гамильтоновых динамических систем. Теорема Лиувилля
дает факт существования квадратур. Поиск оставшейся половины интегралов и переменных разделения (это наши переменные $\gamma_k^{}$) является предметом теории разделения переменных Гамильтона--Якоби. $\Theta$-функции реализуют представления
для динамических переменых $\{p,q\}$ в терминах $\Theta$-рядов.
Более того, изложенный механизм  дает ответ
на вопрос Мозера:
``\ldots\ {\em могут ли все интегрируемые гамильтоновы системы быть описаны с помощью изоспектральной деформации. При этом проблема отыскания интегралов, при условии их существования, сводится к нахождению линейного оператора, чей спектр сохраняется\/}''. Если интегрируемость принадлежит к гиперэллиптической иррациональности (что часто возможно определить отдельно), 
то таким оператором, как видим, всегда может служить оператор Шредингера (1).
Процедура интегрирования остается в рамках теории разделения переменных и поиску канонических (и не канонических) преобразований.

В современных теориях интегрирования линейных дифференциальных уравнений, как это пошло от работ Лиувилля, следует строго определять классы функций на которых задано дифференциальное уравнение. Понятие <<проинтегрировать>> при этом означает операции неопределенного интегрирования над этими объектами + <<естественные>> операции над ними типа алгебры, взятие радикалов, обращения и т.\,д.
В изложенной формулировке теории, как видно, все эти объекты явно предъявляются.

\subsection{Теория солитонов}
\noindent
В заключении скажем что все известные решения и потенциалы
выражающиеся в элементарных функциях (солитоны и их вырождения)
с необходимостью вкладываются в описанную схему. Это происходит 
тогда и только тогда когда голоморфные абелевы интегралы \reff{Jacobi} вычисляются в элементарных функциях, т.\,е. в логарифмах и рациональных функциях.
Как известно это возможно только если в гиперэллиптической кривой \reff{hyper}, а точнее в полиноме $(z-E_1)\cdots(z-E_{2g+1})$, 
несколько корней $E_k$ сливаются так что все абелевы интегралы
редуцируются к иррациональности вида $\sqrt{(z-e_1^{})(z-e_2^{})}$.  В таких случаях говорят, что род $g$ гиперэллиптической кривой \reff{hyper} вырождается в ноль, а интегралы вычисляются в арксинусах, т.\,е. снова сводятся к логарифмам. Обращения таких интегралов
дают экспоненты, которые хорошо известны в теории солитонов.
Кроме солитонов и их разновидностей, решений в элементарных аналитических функциях, соответствующих {\em конечному\/} роду, быть {\em не может\/}. Хотя алгебраическая кривая вырождается в квадратный корень от полинома второй степени, уравнения Новикова по прежнему остаются дифференциальными уравнениями 
порядка $2g+1$, но с решениями в элементарных функциях
(потенциалы Баргмана).

Разумеется построения решений в элементарных функциях путем обращения элементарных абелевых интегралов далеко не самый удобный способ. Для этой цели в теории солитонов существуют самостоятельные прямые методы: преобразования Дарбу--Бэклунда и другие методы <<одевания>> известные в {\sc мозр}.

{

}

\pagebreak
\parindent=1em
\newcommand{\eq}[2]
{{}\\[-1mm]{}\\#1 \hfil $\displaystyle #2\,\,$ \ \ \ \ \hfil\\{}\\[-1mm]}
\hfill {\small\sc Приложение}

\vskip 1cm

\small
\centerline{{\em Comptes Rendus de l'Acad\'emie des Sciences\/}
(1919) {\bf 168}, 47--50.}
\vskip 0.5cm
\centerline{{\sc математический анализ.} ---
{\em Определение случаев редукции}}
\centerline{{\em дифференциального уравнения\/}
$\displaystyle \frac{d^2 y}{dx^2}=[\varphi(x)+h]\,y$.}

\vskip 0.5cm

\centerline{{\bf {\normalsize J}\scriptsize ULES \ {\normalsize D}RACH}}

\vskip 1cm

1. Уравнение
\eq{(1)}{\frac{d^2 y}{dx^2}=[\varphi(x)+h]\,y,}
в котором $\varphi(x)$ обозначает произвольную функцию  $x$ и параметра
$h$, выступает в частности в определении <<гармонических>> решений
<<гармонического>> уравнения
\eq{($a$)}{\frac{\partial^2 Z}{\partial u \,\partial v}= [\varphi(u+v)-
\psi(u-v)]\,Z,}
которое связано в Геометрии с бесконечно малой деформацией
минимальных поверхностей. Впрочем оно является {\em редуцированной формой\/}
общего уравнения
\eq{(2)}{\frac{d}{dt}\!\left( k\,\frac{dV}{dt}\right)+(g\,h-l)\,V =0,}
где $k,\,g,\,l$ --- любые функции $t$, которое мы встречаем
в классических задачах математической физики (охлаждение однородного стержня,
теории Штурма и Лиувилля, и т.\,д.); (2) сводится к (1) двумя квадратурами.

Следовательно, очень важно знать случаи, где упрощение встречается
в интегрировании (1), {\em оставляя параметр $h$ произвольным\/}\,%
\setcounter{footnote}{0}%
(\,\footnote{\hspace{-3.3mm}(\hspace{2.5mm})
\scriptsize Это исследование может быть сделано с помощью
классической теории Э.\,Пикара для линейных уравнений.}\hspace{0.1em}).
Нам удалось определить функцию $\varphi$ во всех случаях, где интеграл
может получаться квадратурами. Мы характеризуем, кроме того, все
другие случаи редукции {\em группы рациональности\/}
уравнения (1); функция $\varphi$ удовлетворяет дифференциальным уравнениям,
которые не могут интегрироваться квадратурами; самый простой случай, снова
дает, например, уравнение Пенлеве:
$$
\frac{d^2\varphi}{dx^2}=3\,\varphi^2+4\,x.
$$

2. Положим $y'=\rho\,y$, резольвента в $\rho$ является уравнением
Риккати
\eq{(3)}{\rho'+\rho^2=\varphi(x)+h,}
и группа рациональности соответствующего уравнения в частных производных,
в области $[h,\,\rho,$ $\varphi(x)]$:
$$
X(f)=\frac{\partial f}{\partial x} +
\frac{\partial f}{\partial \rho}\,(\varphi+h-\rho^2)=0,
$$
образована преобразованиями
$$
h_1^{}=h, \qquad f_1^{}=\frac{a\,f+b}{c\,f+d},
$$
где $a,\,b,\,c,\,d$ являются функциями $h$.

Когда эта группа редуцируется, она становится линейной по $f$ и выражение
инварианта $J=\mfrac{\partial^2f}{\partial\rho^2}:
\mfrac{\partial f}{\partial \rho}$
является рациональным по $\rho$ и $h$.

Корни знаменателя $J$, рассматриваемого как полином по $\rho$,
являются частными решениями (3), алгебраическими по $h$.

Случай одного корня (рационального по $h$)  не возможен.
Также, число корней не может превосходить 2, т.\,е. что интеграл (3) никогда
не является алгебраическим по $h$. Остается исследовать случай двух
частных решений, алгебраических по $h$, для уравнения (3). Они могут
определяться формулами
$$
2\,\rho_1^{}=\frac{R'}{R}+\frac{\sqrt{\Omega}}{R}
\qquad
2\,\rho_1^{}=\frac{R'}{R}-\frac{\sqrt{\Omega}}{R},
$$
где $\Omega$ есть полином по $h$ с постоянными коэффициентами и где полином
по $h$, обозначаемый $R$, удовлетворяет дифференциальному уравнению
\eq{(4)}{R'''-4\,R'\,(\varphi+h)-2\,R\,\varphi'=0;}
Уравнение (3) интегрируется тогда {\em одной квадратурой\/}.

\scriptsize

Уравнение (4) может допускать в качестве решения $R$ полином по $h$
{\em любой степени $n$\/}; тогда получаем для $\varphi$ уравнение порядка
$(2\,n+1)$.  Интегралы, завися от $(n+1)$ констант, получаются если положим

$$
R'^2-2\,R\,R''+4\,R^2\,(\varphi+h)=\Omega(h),
$$

\noindent где полином в правой части --- с постоянными коэффициентами.
Они выражают то, что интеграл $f$ уравнения (3), дробь первой степени по $\rho$,
определяется, с точностью до множителя, функцией $h$.

\small
3. Дарбу заметил\,%
\setcounter{footnote}{0}%
(\,\footnote{\hspace{-3.3mm}(\hspace{2.5mm})
\scriptsize{\em Le\c{c}ons sur la th\'eorie des surfaces\/}, t.\,2, p.\,196,
\S\,408.}\hspace{0.1em}),
что можно перейти от уравнения (1) к другому преобразованием
$$
z=y'-y\,\frac{y'_{0}}{y_0^{}},
$$
где $y_0^{}$ удовлетворяет
$$
y''_0=(\varphi+h_0^{})\,y_0^{}.
$$
Преобразование, которое из этого следует для уравнения Риккати (3) есть
$$
r=\frac{h-h_0^{}}{\rho-\rho_0^{}}-\rho_0^{}
$$
и заменяет $\varphi$ на $\psi=2\,(\rho_0^2-h_0^{})-\varphi$;
оно зависит от двух констант. Но этого преобразования недостаточно,
чтобы перейти от случая, когда $R$ --- степени $n$, к случаю, когда
$R$ --- степени
$(n+1$); новое уравнение порядка $(2\,n+3)$ для $\varphi$ содержит на самом
деле произвольную константу.

4. Другой общий случай редукции уравнения (3) не приводит к его
полному интегрированию. Проективные преобразования интеграла $f$ являются
тогда независимыми от $h$.

Мы здесь имеем, из  (3)
$$
\frac{\partial\rho}{\partial h}= -\frac LP\,\rho^2+\frac MP\,\rho+\frac NP,
$$
$L,\,M,\,N,\,P$ являются полиномами по $h$; последний ---
с постоянными коэффициентами и {\em произволен\/}.
Полином $L$ удовлетворяет уравнению
$$
-L'''+4\,L'\,(\varphi+h)+2\,L\,\varphi'=2\,P,
$$
и, {\em когда $P$ фиксировано\/}, может быть какой-угодно степени.

Преобразование Дарбу еще применяется, но уравнения, которые
получаются  для $\varphi$, определяют новые трансценденты и не
интегрируются квадратурами.

Развитие этих результатов будет являться предметом следующей работы.

\vfill
\bigskip\bigskip\bigskip

\centerline{{\em Comptes Rendus de l'Acad\'emie des Sciences\/}
(1919) {\bf 168}, 337--340.}
\vskip 0.5cm
\centerline{{\sc математический анализ.} ---
{\em Об интегрировании квадратурами}}
\centerline{{\em уравнения\/}
$\displaystyle \frac{d^2 y}{dx^2}=[\varphi(x)+h]\,y$.}

\vskip 0.5cm
\centerline{{\bf {\normalsize J}\scriptsize ULES \ {\normalsize D}RACH}}

\vskip 1cm

\setcounter{footnote}{0}
\noindent
Недавно мы указали\,(\,\footnote{\hspace{-3.3mm}(\hspace{2.5mm})
\scriptsize {\it Comptes Rendus}, t.\,168, 1919, p.\,47.}\hspace{0.1em}),
каковы основные случаи редукции группы
рациональности уравнения
\eq{(H)}{\frac{d^2 y}{dx^2}=[\varphi(x)+h]\,y,}
где $h$ --- {\em произвольный параметр\/}. Самым интересными из них являются
те, в которых уравнение Риккати
$$
\rho'+\rho^2=\varphi+h
$$
[а следовательно уравнение (H)] интегрируется {\em квадратурами\/};
мы сейчас покажем как определяют функцию $\varphi$ во всех этих случаях.

1. Уравнение Риккати допускает здесь два решения
$\mfrac{R'\pm\sqrt{\Omega}}{2\,R}$, где $R$ это полином по $h$
степени $n$
$$
R=h^n+R_1^{}\,h^{n-1}+\ldots
$$
и $\Omega$ --- полином по $h$ степени $(2\,n+1)$ с
{\em постоянными коэффициентами\/}.

Функция $R$ удовлетворяет уравнению третьего порядка
\eq{(A)}{R'''-4\,R'\,(\varphi+h)-2\,R\,\varphi'=0,}
что дает, чтобы определить $\varphi$, уравнение $E_{2 n+1}$
порядка $(2\,n+1)$, зависящее от $n$ произвольных констант
$c_1^{},\,\ldots,\, c_n^{}$.
Именно это уравнение нам удалось проинтегрировать.

Замечая, что (A) эквивалентно своему сопряженному, выводим отсюда первый
квадратичный интеграл%
\eq{(B)}{R'^2-2\,R\,R''+4\,R^2\,(\varphi+h)=\Omega.}
$n$ первых коэффициентов $\Omega$ являются функциями $c_1^{},\,\ldots,\,
c_n^{}$;
$(n+1)$ последних $d_1^{},\,\ldots,\, d_{n+1}^{}$ являются произвольными и
представляют
столько же интегралов $E_{2 n+1}$, {\em целых\/} по
$\varphi,\, \varphi', \,\ldots$, $\varphi^{(2 n+1)}$. Остается, следовательно,
проинтегрировать  уравнение порядка $n$ с $(2\,n+1)$ константами.

2. Положим
$$
R=h^n+R_1^{}\,h^{n-1}+\ldots =
(h-\omega_1^{})\,(h-\omega_2^{})\,\ldots \,(h-\omega_n^{});
$$
$\omega_i^{}$ являются алгебраическими функциями от
$\varphi,\, \varphi', \,\ldots, \,\varphi^{(2 n-2)}$ и мы будем иметь
$$
-\frac{R'}{R}=\frac{\omega_1'}{h-\omega_1^{}} +\ldots
+\frac{\omega_n'}{h-\omega_n^{}}.
$$

Замечая, что корни $h=\omega_i^{}$ аннулируют $R'^2-\Omega$ вследствие (B),
т.е. один из
факторов $R'+\sqrt{\Omega},\,\,R'-\sqrt{\Omega}$, можно будет записать
$n$ уравнений
$$
\frac{\varepsilon_i^{}\,\omega_i'}{\sqrt{\Omega_i}}=
\frac{1}{(\omega_i^{}-\omega_1^{})\,\ldots\, (\omega_i^{}-\omega_n^{})}
\qquad (i=1,\,\ldots, \,n),
$$
где $\Omega_i=\Omega(\omega_i^{}),\,\, \varepsilon_i^{}=\pm 1.$

Выводим отсюда искомые $n$ интегралов $u_i^{}$ в виде
$$
\int_{\alpha_1^{}}^{\omega_1}\!
\frac{\omega_1^{\lambda}\,d\omega_1^{}}{\sqrt{\Omega_1}}
+\ldots +
\int_{\alpha_n^{}}^{\omega_n}\!\frac{\omega_n^{\lambda}\,d\omega_n^{}}
{\sqrt{\Omega_n}}=u_{\lambda}^{} \qquad (\lambda=0,\,1,\,\ldots, \,n-2),
$$
$$
\int_{\alpha_1^{}}^{\omega_1}\!
\frac{\omega_1^{n-1}\,d\omega_1^{}}{\sqrt{\Omega_1}}
+\ldots +
\int_{\alpha_n^{}}^{\omega_n}\!\frac{\omega_n^{n-1}\,d\omega_n^{}}
{\sqrt{\Omega_n}}=x+u_{n-1}^{},
$$
при условии, что пути интегрирования
при $\varepsilon_i^{}=-1$ устанавливаются таким образом, что
прямолинейный интеграл $I$  заменяется на $2\,A-I$, где $A$
обозначает подходящую константу; это
дает $\Big(\!\mfrac n2\!\Big)$ различных по форме систем.

3. Функция $\varphi$, которая задана как
$$
\omega_1^{}+\omega_2^{}+ \ldots + \omega_n^{}=
-R_1^{}=-C_1^{}+\mfrac12\,\varphi,
$$
и элементарные симметрические функции  $\omega$ \big[так же как и
симметрические функции $\sqrt{\Omega_i}\,$\big] являются
{\em абелевыми функциями\/}
(гиперэллиптическими) аргументов $u_0^{},\,\ldots, \,u_{n-1}^{}+x$,
т.е. {\em однозначными\/}, {\em мероморфными\/}
этих элементов, обладающие  $2\,n$ системами
периодов\,%
\setcounter{footnote}{0}%
(\,\footnote{\hspace{-3.3mm}(\hspace{2.5mm}) \scriptsize{\sc Weierstrass},
{\em Journal de Crelle\/}, t.\,47 et 52.}\hspace{0.1em}).
Функция $\varphi$  от переменной $x$ в общем не
является периодической, но она вновь обретает свое значение
когда добавляют
одновременно к $x,\,u_0^{},\,\ldots,\,u_{n-2}$ соответствующие периоды:
это определяет {\em группу монодромии\/} интегралов $u_i^{}$. Случай
$n=1$  дает для $\varphi$ обратную функцию эллиптического интеграла
первого рода.

Аналогичное рассуждение, но относящееся к другим корням
$\xi_1^{},\,\ldots,\,\xi_{n+1}$ уравнения $R'^2-\Omega=0$, дало бы
$(n+1)$ трансцендентных интегралов $E_{2n+1}$; теорема Абеля легко
показывает, что эти интегралы  выражаются через предыдущие и
алгебраические интегралы $d_1^{},\,\ldots,\,d_{n+1}^{}$.

Фундаментальными интегралами уравнения (H) являются
$\displaystyle\sqrt{R\,}\,e_{\mathstrut}^{{\frac12\!
{\displaystyle \sqrt{\mbox{\footnotesize $\Omega\,$}}\!
\scalebox{0.7}[0.7]{\raisebox{0.2em}
{\mbox{\footnotesize$\displaystyle\int$}}}}\!
\textstyle\frac{dx}{R}}}$;
заметим, что мы имеем
$$
\sqrt{\Omega}\int \!\frac{dx}{R}=
\sum\,\int_{\alpha_i^{}}^{\omega_i^{}}\!\displaystyle\frac{\sqrt{\mathstrut
\Omega(h)}}{(h-\omega_i^{})}
\,\frac{d\omega_i^{}}{\sqrt{\Omega(\omega_i^{})}}.
$$

Эта сумма интегралов третьего рода является суммой логарифма абелевой функции
и абелевой функции. Преобразования, которые претерпевают интегралы (H)
выводятся отсюда.

Добавим, что уравнение (H) допускает
преобразование {\em в себя\/}
$$
Y=[\theta(h)-R']\,y+2\,R\,y',
$$
где $\theta(h)$ --- полином по $h$ с постоянными коэффициентами.

4. Изучение уравнений (H), которые интегрируется квадратурами является
следовательно не чем иным как изучением {\em абелевых функций\/} и их
{\em вырождений\/}, когда $\Omega$
имеет кратные корни (исследование сделанное в деталях Эмилем Пикаром
и Пенлеве для $n=2$).

Преобразование Дарбу, которое замещает  $\varphi$ на
$2\,(\rho_0^{2}-h_0^{})-\varphi$, где $\rho_0'+\rho_0^2=\varphi+h_0^{}$,
и $R$ на полином степени $(n+1)$ по $h$, не приводит к
{\em собственным\/} трансцендентам  индекса $(n+1)$;
оно умножает $\Omega$ на $(h-h_0^{})^2$, что не
меняет {\em жанр\/}. Наоборот, если $\Omega$ содержит множитель $(h-h_0^{})^2$,
преобразование Дарбу относительно $h_0^{}$ и
$\mfrac{R'(h_0^{})}{2R(h_0^{})}$ понижает на единицу
индекс  $n$.
Функции $\varphi$, {\em несобственные\/} для индекса $n$,
получаются
суперпозицией квадратур, начиная с функций, {\em собственных\/} для низшего
индекса; они однозначные, но  {\em не мероморфные\/}.

Наконец для значений $h=h_i^{}$ которые аннулируют $\Omega$, мы имеем уравнение
$y''=(\varphi+h_i^{})\,y$, которое допускает {\em абелево\/} решение
$\sqrt{{}^{\mathstrut} R(h_i^{})}$, следовательно, однозначное по
$u_0^{}, \,\ldots, \,u_{n-1}^{}+x$.

Здесь мы даем естественное развитие важных исследований
Эрмита и Эмиля Пикара  об уравнении Ла\-ме\,%
\setcounter{footnote}{0}%
(\,\footnote{\hspace{-3.3mm}(\hspace{2.5mm}) \scriptsize{\em Comptes Rendus\/},
t.\,85, 1877,
p.\,689, et t.\,89, 21 juillet 1879.}\hspace{0.1em}),
для которого $\varphi=n\,(n+1)\,k^2\,\mbox{sn}^2x$;
и последующие работы Бриоски, Эллиота, Фукса, Дарбу по аналогичным уравнениям.
Заметим, что когда $\varphi$ выбирается как независимая переменная,
$\mfrac{d\varphi}{dx}$ является {\em однозначной\/} по $x$, но
{\em трансцендентной\/} по $\varphi$, за исключением случая когда функции
$\varphi$ являются производными от эллиптических функций,
единственный случай, рассмотренный до настоящего времени.

\end{document}